\documentclass[11pt,twoside]{article}
\usepackage{asp2010}

\resetcounters

\begin{document}

\title{Direct Evidence for Termination of Obscured Star Formation by Radiatively Driven Outflows in FeLoBAL QSOs}
\author{Duncan Farrah,$^1$ Tanya Urrutia,$^2$ Mark Lacy,$^3$ Andreas Efstathiou,$^4$ Jose Afonso,$^5$ Kristen Coppin,$^6$ Patrick B. Hall,$^7$ Carol Lonsdale,$^3$ Tom Jarrett,$^8$ Carrie Bridge,$^8$ Colin Borys,$^8$ and Sara Petty$^9$
\affil{$^1$University of Sussex, Falmer, Brighton BN1 9QH, UK}
\affil{$^2$Leibniz Institute for Astrophysics, Potsdam, Germany}
\affil{$^3$NRAO, Charlottesville, VA 22903, USA}
\affil{$^4$School of Sciences, European University Cyprus, Nicosia, Cyprus}
\affil{$^5$Observat\'{o}rio Astron\'{o}mico de Lisboa, Universidade de Lisboa, Lisbon, Portugal}
\affil{$^6$Department of Physics, McGill University, Montr\'{e}al, Qu\'{e}bec, Canada}
\affil{$^7$Department of Physics \& Astronomy, York University, Toronto, Canada}
\affil{$^8$California Institute of Technology, Pasadena, CA 91125, USA}
\affil{$^9$UCLA, Physics and Astronomy Building, Los Angeles, CA 90095, USA}}

\begin{abstract}
We use SDSS spectra and optical to far-infrared photometry for a sample of 31 FeLoBAL QSOs to study the relationship between the AGN-driven outflows, and obscured star formation in their host galaxies. We find that FeLoBAL QSOs invariably have IR luminosities exceeding $10^{12}$L$_{\odot}$. The AGN supplies 75\% of the total IR emission, on average, but with a range from 20\% to 100$\%$. We find a clear anticorrelation between the strength of the AGN-driven outflows and the contribution from star formation to the total IR luminosity, with a much higher chance of seeing a starburst contribution in excess of 25\% in systems with weak outflows than in systems with strong outflows. Moreover, we find no evidence that this effect is driven by the IR luminosity of the AGN. We conclude that radiatively driven outflows from AGN act to curtail obscured star formation in the host galaxies of reddened QSOs to less than $\sim 25\%$ of the total IR luminosity. This is the most direct evidence yet obtained for `quasar mode' AGN feedback. 
\end{abstract}

\section{Introduction}
Over the last decade, several problems have arisen in modelling the cosmological evolution of galaxies. These include (1) the difficulties that models faced in explaining the observed galaxy luminosity function at low and high redshifts simultaneously (e.g. \citealt{benson03}), (2) the prediction that rich galaxy clusters should harbour cooling flows when few are observed (e.g. \citealt{peterson01}), and (3) the difficulties that models face in reproducing the number of IR-luminous galaxies observed at high redshift (e.g. \citealt{baugh05}).

One of the most promising solutions to these issues is AGN feedback. AGN feedback is the exertion of influence of an SMBH on $\gtrsim$kpc scales to curtail star formation in the host galaxy, and/or accretion onto the SMBH itself. Galaxy evolution models usually assume this feedback to occur in one or both of two simplified modes; `quasar' mode and `radio' mode. Quasar mode feedback occurs via radiation from an accretion disk, while radio mode feedback occurs via a relativistic jet that transfers momentum to the ISM. Both these feedback paradigms have led to improvements in the ability of models to reproduce observations (e.g. \citealt{bower08,somer08}), but observational evidence for either paradigm remains sparse. 

Our group has been looking for evidence of quasar mode feedback by examining systems in which such feedback may be ongoing. To do so, we have been studying the `FeLoBAL' QSOs \citep{hazard87,hall02}. We selected this population for two reasons. First, their UV absorption troughs are unambiguous signatures of radiatively driven outflows powered by an AGN. Second, FeLoBAL QSOs are invariably reddened and have high IR luminosities \citep{farrah07,farrah10}. We here present results for 31 FeLoBAL QSOs, in which we compare the strength of their outflows as estimated from their UV spectral properties, to the luminosity of obscured star formation in their host galaxies as measured from optical through far-IR photometry. We define "IR luminosity" as the luminosity integrated over 1-1000$\mu$m in the rest-frame.

\section{Methods}
We assembled our sample from the SDSS. We started with the six SDSS objects in \citealt{farrah07} (hereafter F07) and added a further 25 FeLoBAL QSOs from \citet{trump06} over the redshift range $0.8<z<1.8$ (chosen so that the SDSS spectra always contain the Mg II BAL). There was no selection on IR luminosity. We observed the sample with MIPS \citep{rieke04} onboard Spitzer using standard parameters. We reduced the MIPS data using the MOPEX software. We then added archival photometry from the SDSS, 2MASS or UKIDSS \citep{skrutskie06,law07} and WISE \citep{wright10} to give photometric coverage from 0.9$\mu$m to 160$\mu$m

We measure total, starburst and AGN IR luminosities by fitting the optical through MIPS photometry for each object with radiative transfer models for AGN and starbursts, following the methods in \citealt{farrah03} and F07. For the AGN models, we follow \citet{efs95} and use a tapered disk dust distribution. For the starburst models we use the templates of \citet{efs00} with the updated dust model of \citet{efs09}. The models do not include  broad absorption features in the rest-frame UV that arise from radiatively driven outflows, so we do not include the SDSS spectra, or photometry shortward of $\sim$0.35$\mu$m in the rest-frame, in the fits. In 27/31 cases the models provide a good fit to the data ($\chi^{2}_{red}<2$). The remaining four objects have $2<\chi^{2}_{red}<3$, usually due to a poor fit in the near-IR. This could be due to the lack of a host galaxy contribution in our models, so we defer exploration of this until more comprehensive near-IR data are available.

\section{Results \& Discussion}
We find that FeLoBAL QSOs are luminous in the IR. All 31 of our sample have total IR luminosities (L$_{Tot}$) in excess of $10^{12}$L$_{\odot}$, with nine objects exceeding $10^{13}$L$_{\odot}$. These luminosities are comparable to those of the wider population of BAL QSOs \citep{gallagher07}, red QSOs \citep{geo09}, and ULIRGs \citep{farrah03,farrah09}. The dominant power source is usually an AGN. A pure AGN is either the most likely power source, or consistent within the 90\% confidence interval, for $35\%$ of the sample. A starburst component is required for the remaining objects, but in only twelve objects is the starburst more luminous than $10^{12}$L$_{\odot}$, and in only three objects is the starburst more luminous than the AGN. The mean starburst contribution to the total IR luminosity (f$_{SB}$) is $\sim24\%$, comparable to that of local ULIRGs with `warm' IR colours, but lower than those of PG QSOs \citep{veill09}. 

\begin{figure}
\includegraphics[width=52mm,angle=90]{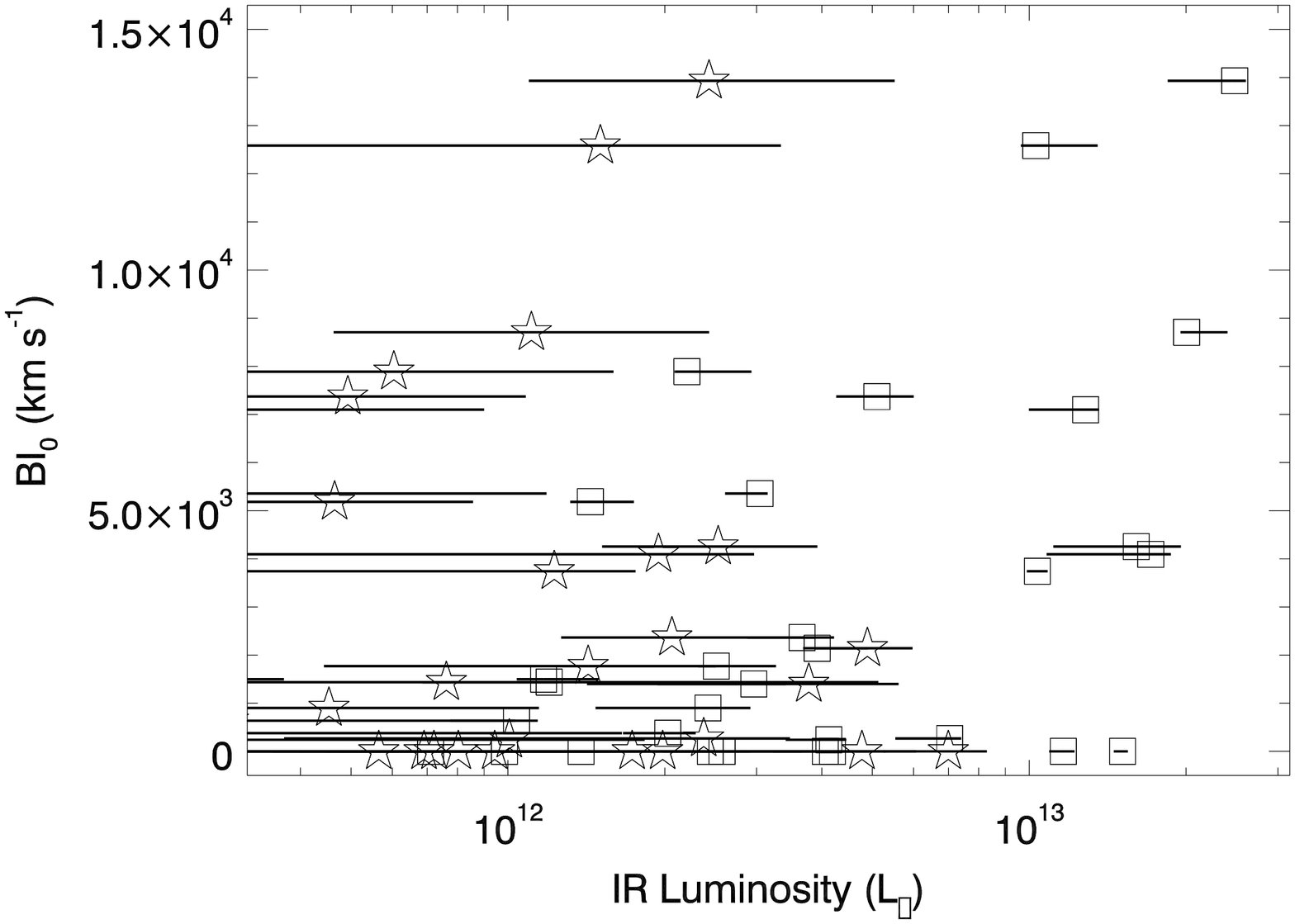}
\includegraphics[width=52mm,angle=90]{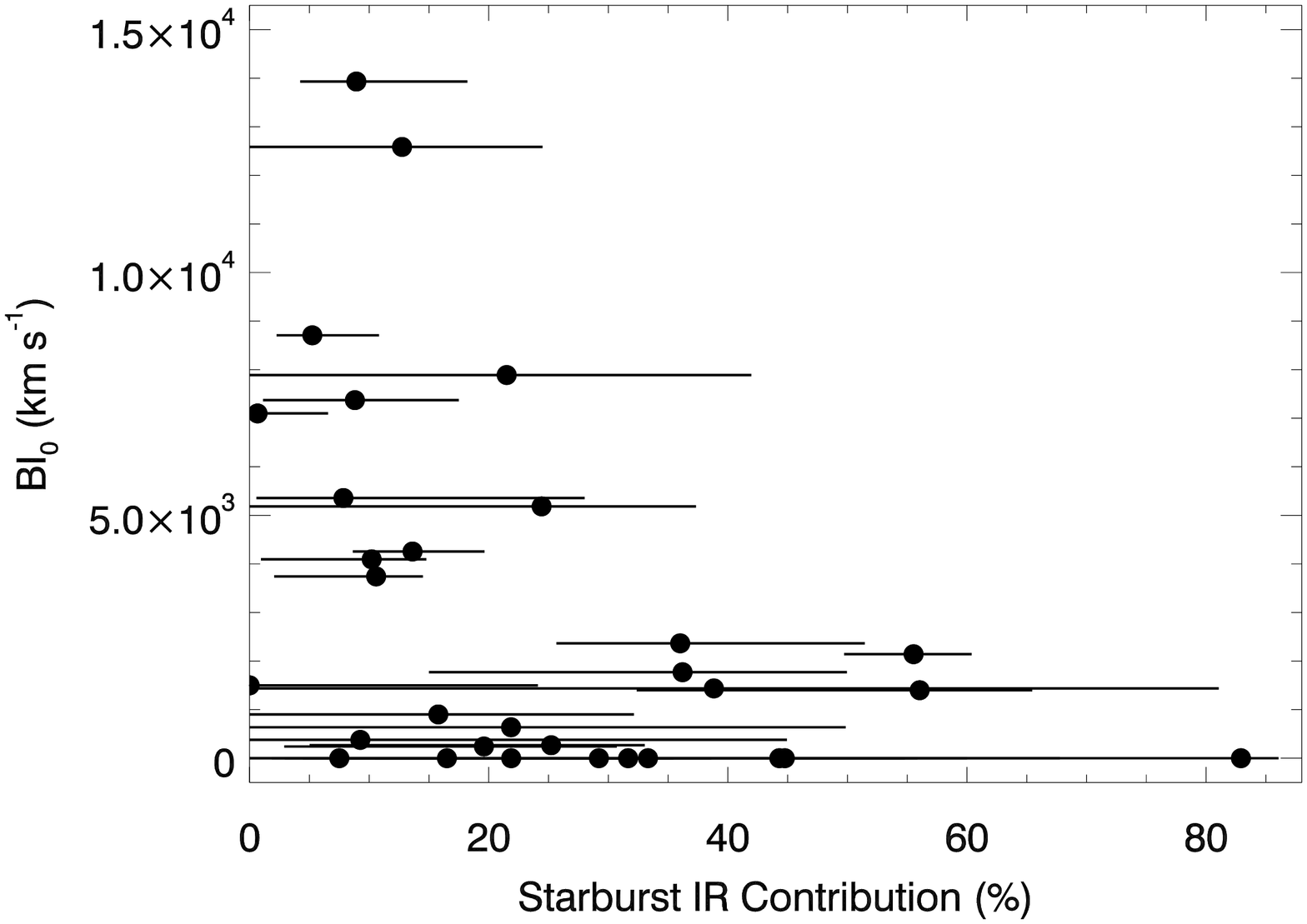}
\caption{Absorption strength vs (left) starburst (stars) and AGN (squares) IR luminosities, and (right) starburst contribution to the total IR luminosity.}\label{luminosities}
\end{figure}

We now examine whether or not there is a relationship between the AGN-driven outflows and the obscured star formation. To quantify the strength of the outflows we use the same, single species (MgII$\lambda$2799\AA) across the whole sample, and adopt the BI$_{0}$ parameter of \citet{gibson09}. We measured the MgII absorption strengths using the methods described in \citet{f2ms}. 

We first examine if absorption strength depends on AGN or starburst IR luminosity (Fig \ref{luminosities}, left). There is no correlation between absorption strength and L$_{SB}$ (a Spearman rank test gives $\rho = -0.10$, $P = 0.58$) and at best a weak correlation between absorption strength and L$_{AGN}$ ($\rho = 0.39$, $P = 0.04$). Conversely, we see a weak but clear anticorrelation between absorption strength and f$_{SB}$ ($\rho = -0.49$, $P = 0.005$, Fig \ref{luminosities} right). Moreover, all the systems with BI$_{0}\gtrsim 3500$km s$^{-1}$ have $f_{SB}<25\%$, while the systems with BI$_{0}\lesssim 3500$km s$^{-1}$ have a wide dispersion in starburst contributions, from $0\%$ to $\sim80\%$. A two-sided Kolmogorov-Smirnov test for the objects above and below BI$_{0} = 3500$km s$^{-1}$ yields a difference in distributions at 99.84\% confidence, though the number of objects in the two subsamples is too low for this test to be robust. 

\begin{figure}
\includegraphics[width=52mm,angle=90]{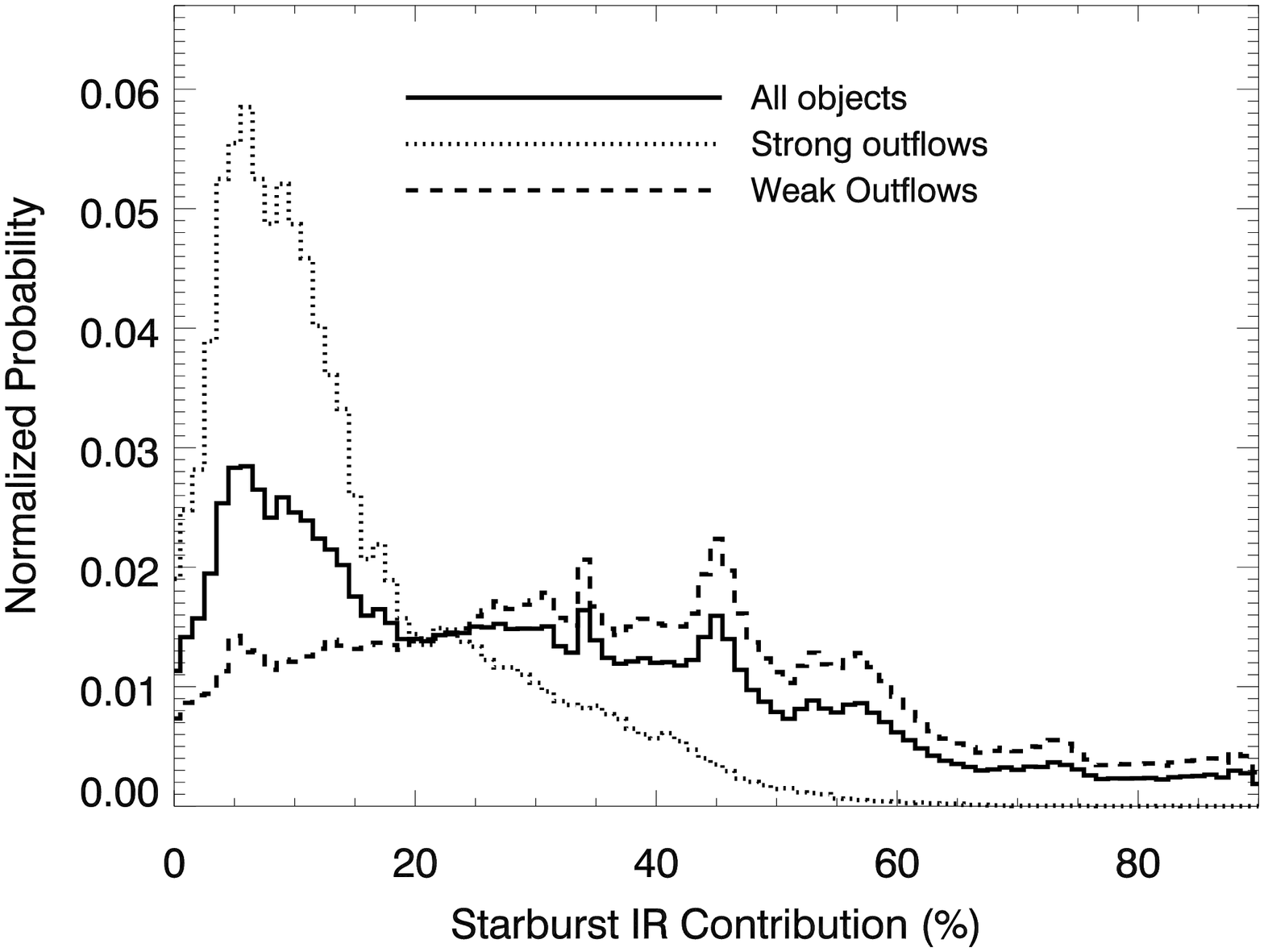}
\includegraphics[width=52mm,angle=90]{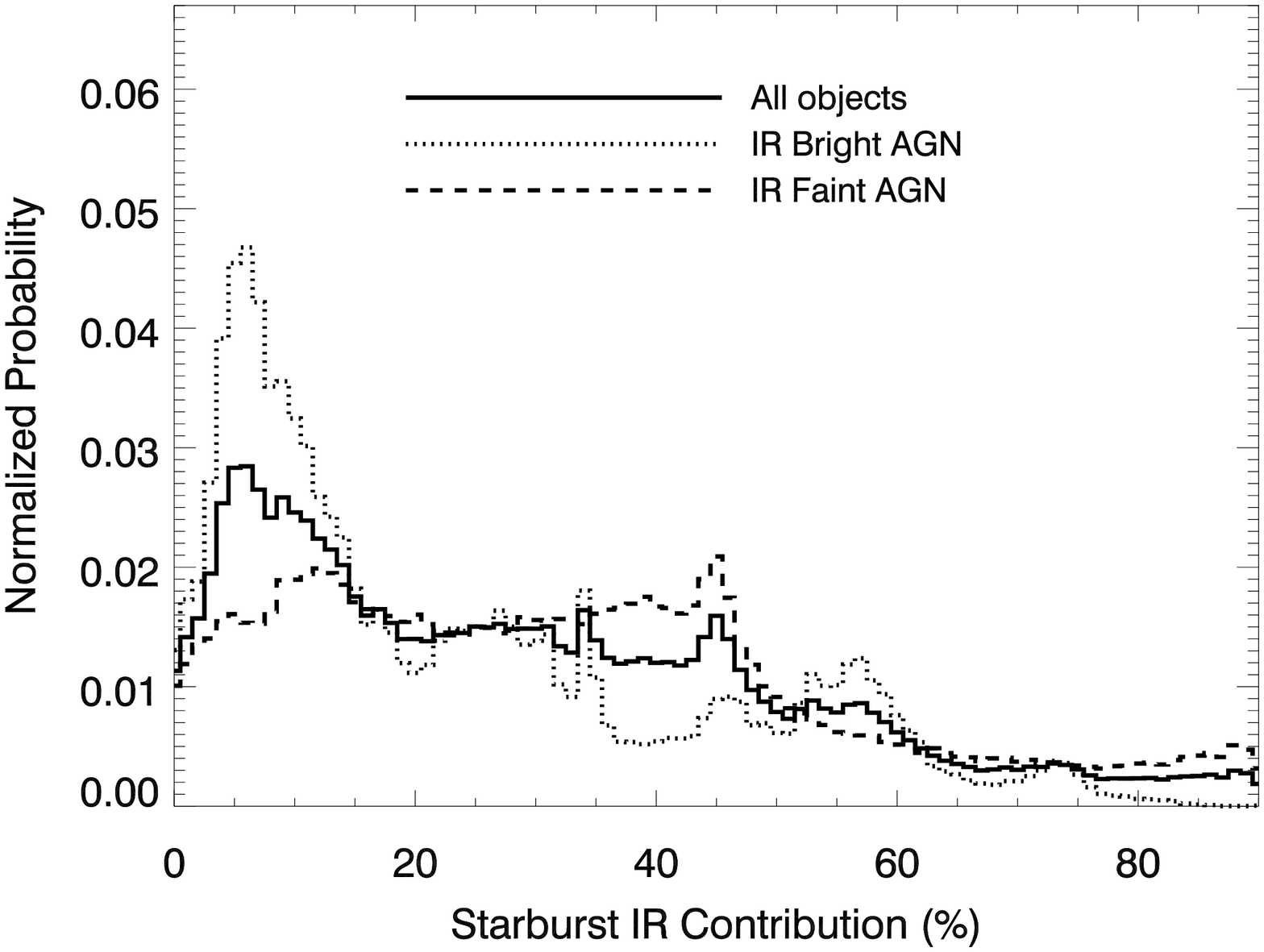}
\caption{Probability Distribution Functions for the starburst contribution to the total IR luminosity. In both panels, the solid line is the PDF for the whole sample. The other lines are for subsamples divided according to two criteria. {\itshape Left panel}: by absorption strength. {\itshape Right panel}: by AGN IR luminosity.}\label{pdfs}
\end{figure}

We examine this result in more detail by constructing probability distribution functions (using all the fit solutions) for f$_{SB}$ for two subsamples divided at BI$_{0}=3500$km s$^{-1}$ (Figure \ref{pdfs}, left). We see a clear difference. The low absorption strength subsample shows a much higher chance of a higher starburst contribution than the high absorption strength subsample. We quantify this by extracting the probabilities of obtaining $f_{SB}>25\%$. For the whole sample we find P($f_{SB}>25\%$)$=50.3^{+5.3}_{-5.4}$\%, for the low absorption strength sample we find P($f_{SB}>25\%$)$=67.3^{+4.5}_{-4.1}$\%, while for the high absorption strength sample we find P($f_{SB}>25\%$)$=17.8^{+3.7}_{-6.5}$\%. 

This anticorrelation between absorption strength and f$_{SB}$ is straightforwardly interpreted as the outflow from the AGN curtailing star formation in the host galaxy. In this scenario, the systems with $f_{SB}>25\%$ are those in which an outflow has yet to curtail star formation, so the outflows always have BI$_{0}\lesssim 3500$km s$^{-1}$. Conversely, the systems with $f_{SB}<25\%$ are those in which an outflow has curtailed star formation, {\itshape and} those in which such an outflow has subsequently waned, making the dispersion in absorption strengths wide.

There are however four other ways that this anticorrelation could arise. The first is that the starburst `suppresses' AGN outflows, so when the starburst wanes (via a cause unrelated to the AGN) an AGN driven outflow can appear, possibly because the ISM density has been reduced by the starburst. We cannot test this scenario, but it would likely require a serendipitous conjunction of ISM and SMBH parameters, so we do not consider it further. The second is if a high starburst contribution meant that the Mg II troughs were {\it observed} to be weaker than they really are. Again, we cannot test this, but the rest-frame UV continua of starbursts in ULIRGs are at least an order of magnitude too weak to provide this effect, and can show absorption in the same species \citep{farrah05}. The third is if BAL QSOs with strong starbursts preferentially drop out of the SDSS QSO selection compared to BAL QSOs with weak starbursts, and were thus not included in \citet{trump06}. This possibility is also not testable by us, but the SDSS is now turning up FeLoBAL QSOs in large numbers, and the SDSS QSO followup colour selections are fairly relaxed, so we do not consider this possibility likely either. 

The fourth possibility is that stronger outflows reflect an increase in the IR emission from the AGN, but have no effect on the starburst; if this is the case we would see a decline in the contribution from star formation to the total IR luminosity with increasing absorption strength, but with no direct relationship behind the decline. This is a possibility we can test, as follows. If outflow strength is just a proxy for AGN luminosity, then we should see a bigger difference between the starburst contribution PDFs for subsamples divided by AGN luminosity than between subsamples divided by absorption strength. In the right panel of Figure \ref{pdfs} we show starburst contribution PDFs for two subsamples divided by AGN luminosity at L$_{AGN}=10^{12.5}$L$_{\odot}$. The difference between the PDFs divided by AGN luminosity is {\it weaker} than the difference between the PDFs divided by absorption strength.  Furthermore, we are more likely (albeit only at $\sim$2$\sigma$) to obtain a smaller starburst contribution by selecting high absorption strength systems than we are by selecting high AGN luminosity systems; for P($f_{SB}>25\%$): the BI$_{0}>3500$km s$^{-1}$ subsample is $17.8^{+3.7}_{-6.5}$\% while the L$_{AGN}>10^{12.5}$L$_{\odot}$ subsample is $38.7^{+8.5}_{-10.0}$\%). In other words, we are more successful in finding systems with a large starburst contribution to the total IR emission by selecting on weak outflows than we are by selecting on low AGN luminosity. Also, we are more likely to find star formation with a lower absolute luminosity in the {\itshape lower} luminosity AGN subsample (e.g. for P(L$_{Sb}<10^{12}$L$_{\odot})$: the L$_{AGN}<10^{12.5}$L$_{\odot}$ subsample is $69.4^{+5.2}_{-4.6}$\% while the L$_{AGN}>10^{12.5}$L$_{\odot}$ subsample is $43.2^{+6.9}_{-10.4}$\%)

Overall therefore, we find that radiatively driven outflows from an AGN with absorption strengths $\gtrsim3500$ km s$^{-1}$ act to curtail star formation in their host galaxies. We also find that this effect is (at least largely) relative; such outflows reduce the contribution from star formation to the total IR luminosity to less than $\sim25\%$. We also propose that the {\itshape infrared} luminosity of the AGN is not a good proxy for the degree of AGN feedback that is taking place. 

%\acknowledgements 


\begin{thebibliography}{}

\bibitem[Baugh et al.(2005)]{baugh05} 
Baugh, C.~M., et al.\ 2005, \mnras, 356, 1191 

\bibitem[Bower et al.(2008)]{bower08} Bower, R.~G., McCarthy, 
I.~G., \& Benson, A.~J.\ 2008, \mnras, 390, 1399 

\bibitem[Benson et al.(2003)]{benson03} 
Benson, A.~J., et al.\ 2003, \apj, 599, 38 

\bibitem[Efstathiou \& Rowan-Robinson(1995)]{efs95} 
Efstathiou, A., \& Rowan-Robinson, M.\ 1995, \mnras, 273, 649 

\bibitem[Efstathiou et al.(2000)]{efs00} Efstathiou, A., 
Rowan-Robinson, M., \& Siebenmorgen, R.\ 2000, \mnras, 313, 734 

\bibitem[Efstathiou \& Siebenmorgen(2009)]{efs09} 
Efstathiou, A., \& Siebenmorgen, R.\ 2009, \aap, 502, 541 

\bibitem[Farrah et al.(2003)]{farrah03} 
Farrah, D., et al.\ 2003, \mnras, 343, 585 

\bibitem[Farrah et al.(2005)]{farrah05} Farrah, D., Surace, 
J.~A., Veilleux, S., Sanders, D.~B., \& Vacca, W.~D.\ 2005, \apj, 626, 70 

\bibitem[Farrah et al.(2007)]{farrah07} Farrah, D.; Lacy, M.; Priddey, R.; 
Borys, C.; Afonso, J. 2007, \apj, 662, 59

\bibitem[Farrah et al.(2009)]{farrah09} Farrah, D., et al.\ 
2009, \apj, 700, 395 

\bibitem[Farrah et al.(2010)]{farrah10} Farrah, D., et al.\ 
2010, \apj, 717, 868 

\bibitem[Gallagher et al.(2007)]{gallagher07} Gallagher, S. C., et al. 2007, \apj, 665, 157

\bibitem[Georgakakis et al.(2009)]{geo09} 
Georgakakis, A., et al.\ 2009, \mnras, 394, 533 

\bibitem[Gibson et al.(2009)]{gibson09} Gibson, R.~R., et al.\ 
2009, \apj, 692, 758 

\bibitem[Hall et al.(2002)]{hall02} Hall, P.~B., et al.\ 2002, 
\apjs, 141, 267 

\bibitem[Hazard et al.(1987)]{hazard87} Hazard, C., McMahon, R. G., Webb, J. 
K., Morton, D. C. 1987, \apj, 323, 263

\bibitem[Lawrence et al.(2007)]{law07} 
Lawrence, A., Warren, S.~J., Almaini, O., et al.\ 2007, \mnras, 379, 1599 

\bibitem[Lynds(1967)]{lyn67} 
Lynds, C.~R.\ 1967, \apj, 147, 396 

\bibitem[Peterson et al.(2003)]{peterson01} 
Peterson, J.~R., Kahn, S.~M., Paerels, F.~B.~S., et al.\ 2003, \apj, 590, 207 

\bibitem[Rieke et al.(2004)]{rieke04} 
Rieke, G.~H., et al.\ 2004, \apjs, 154, 25

\bibitem[Skrutskie et al.(2006)]{skrutskie06} Skrutskie, M.~F., et al.\ 2006, \aj, 131, 1163 

\bibitem[Somerville et al.(2008)]{somer08} 
Somerville, R.~S., et al.\ 2008, \mnras, 391, 481 

\bibitem[Trump et al.(2006)]{trump06} Trump, J.~R., et al.\ 
2006, \apjs, 165, 1 

\bibitem[Urrutia et al.(2009)]{f2ms} Urrutia, T. et al. 2009, \apj, 698, 1095

\bibitem[Veilleux et al.(2009)]{veill09} Veilleux, S., et al.\ 
2009, \apjs, 182, 628 

\bibitem[Wright et al.(2010)]{wright10} Wright, E.~L., et al. 2010, \aj, 140, 1868 

\end{thebibliography}
\end{document}